\documentstyle[12pt]{article}

\begin{document}
\begin{titlepage}
\setcounter{page}{1}
\title{Multi-black-holes in three dimensions}
\author{G\'erard Cl\'ement \\
\small Laboratoire de Gravitation et Cosmologie Relativistes \\
\small Universit\'e Pierre et Marie Curie, CNRS/URA769 \\
\small Tour 22-12, Bo\^{\i}te 142 \\
\small 4 place Jussieu, 75252 Paris cedex 05, France}
\maketitle
\begin{abstract}
We construct time-dependent multi-centre solutions to three-
\linebreak
dimensional general relativity with zero or negative cosmological
\linebreak
constant. These
solutions correspond to dynamical systems of freely falling black holes
and conical singularities, with a multiply connected spacetime topology.
Stationary multi-black-hole solutions are possible only in the extreme
black hole case.
\end{abstract}
\end{titlepage}

In a now well-known paper \cite{r1}, Ba\~nados et al. gave a black hole
solution
to the three-dimensional Einstein equations with negative cosmological constant
$\Lambda=-l^{-2}$,
and studied its properties. This regular solution, which has
inspired a number of recent papers \cite{r2}, is given by
\begin{eqnarray} \label{1}
ds^2=\nu^2[\frac{r^2}{l^2}-M+\frac{J^2}{4r^2}]dt^2&-&r^2[d\theta-
\frac{\displaystyle\nu J}{\displaystyle 2r^2}dt]^2
\nonumber \\
&-&\frac{\displaystyle l^2dr^2}{\displaystyle r^2-Ml^2
+\frac{\displaystyle J^2l^2}
{\displaystyle 4r^2}}\,,
\end{eqnarray}
where $\theta$ is periodic with period $2\pi$, the two parameters $M$ and $J$
(with $M\geq0$, $|J|\leq ML$)
are interpreted \cite{r1} as the mass and angular momentum of the black hole,
and the constant $\nu$ ($\equiv N(\infty)$ in \cite{r1}) sets the scale of
time.

     The purpose of the present work is to construct exact dynamical
multi-black-hole
solutions to three-dimensional cosmological gravity. Some time ago,
conformal techniques were used to construct static \cite{r3} and stationary
\cite{r4} multi-center
solutions to pure gravity ($\Lambda = 0$) associated with configurations of
massive and spinning point particles, as well as a class of static
multi-particle solutions to cosmological gravity ($\Lambda\neq{0}$) \cite{r5}.
Using similar methods, we
shall construct what at first sight appear to be stationary multi-black-hole
solutions. However these stationary solutions turn out to be unconsistent,
being generically plagued by extra conical singularities, which are unphysical
in the sense that their worldlines are not geodesics of the multi-black-hole
space-time \cite{r5}. As we shall show, by taking the positions of the
black-hole centers to be no longer constant but time-dependent, one can derive
intrinsically
dynamical solutions corresponding to systems of freely falling black holes
together with ---now physical--- auxiliary conical singularities.

     We first consider for simplicity the case of pure gravity which also
admits a solution which, under certain circumstances, behaves as
a black hole solution. This may be obtained from the $J = 0$
black hole of eq.(\ref{1}) by putting
$\nu={\gamma l^2}/c$, $M=c^2l^{-2}$, $r=c/{cos(l^{-1}X)}$,
$\theta=c^{-1}Y$, leading to
\begin{equation} \label{2}
ds^2 = \gamma^2 l^2 \tan^2(l^{-1}X)dt^2 - \frac{1}{\cos^2(l^{-1}X)}(dX^2+dY^2),
\end{equation}
and taking the limit $l\rightarrow\infty$, which yields
\begin{equation} \label{3}
ds^2 = \gamma^2X^2dt^2 - dX^2 - dY^2 .
\end{equation}
We recognize in (\ref{3}) the well-known two-dimensional Rindler
space-time \cite{r6}
with an extra compact spatial dimension $Y$. As discussed in \cite{r6},
the transformation $\bar{t}=X\sinh(\gamma t)$, $\bar{x}=X\cosh(\gamma t)$,
$\bar{y}=Y$ maps the metric (\ref{3}) into the two disjoint regions $I$
($\bar{x}^2>\bar{t}^2$) of the Minkowski cylinder
$ds^2=d\bar{t}^2-d\bar{x}^2-d\bar{y}^2$
(with $\bar{y}$ periodic). The remaining two regions $II$ ($\bar{t}^2>
\bar{x}^2$) of the Minkowski cylinder may be obtained by extending the
metric (\ref{3}) through the horizon $X^2=0$ (of perimeter $2\pi c$) to
$X^2=-\hat{X}^2<0$, and making the transformation
$\bar{t}=\hat{X}\cosh(\gamma t)$, $\bar{x}=\hat{X}\sinh(\gamma t)$,
$\bar{y}=Y$. The resulting Penrose diagram is shown in Fig. 1. Of course,
this maximally extended Rindler cylinder is undistinguishable from the
Minkowski cylinder. The distinction comes about if for instance the
metric (\ref{3}) arises as an interior solution generated by a ring of
exotic matter \cite{r7}: then the region inside the ring has its own points
at timelike infinity, distinct from those of the exterior region
(Fig. 2), and is thus a genuine black hole (this fact was not fully
appreciated in \cite{r7}). Let us here mention that the four-dimensional
Rindler cylinder, $ds^2=\gamma^2X^2dt^2-dX^2-dY^2-dZ^2$ with $Y$ periodic,
may similarly arise as an interior solution generated by an infinite cylinder
of exotic matter, leading to a black cosmic string \cite{r8}.

To construct multi-black-hole solutions to pure three-dimensional gravity,
we recall that the conformal map $X+iY=Z(z)$ (with $z=x+iy$) generates from
(\ref{3}) the family of stationary flat metrics \cite{r4}
\begin{equation} \label{4}
ds^2=\gamma^2X^2(z,\bar{z})dt^2 - |Z'(z)|^2dzd\bar{z}
\end{equation}
($Z'\equiv dZ/dz$). Consider the multi-center map
\begin{equation} \label{5}
Z=\sum_{i=1}^n c_i\ln(z-a_i)+d
\end{equation}
($c_i$ and $d$ real, $a_i$ complex) of the region $X(z,\bar{z})>0$ of the
Euclidean
$(x,y)$ plane into the spatial sections of the three-dimensional Rindler
space-time; this map preserves spatial infinity ($X\rightarrow+\infty\:
\Leftrightarrow\:|z|\rightarrow+\infty$) if all the $c_i$ are positive.
For $n=1$ and $a_1=0$, $c_1=c>0$, we recover the Rindler cylinder with
$X=c\ln r+d$, $Y=c\,\theta$, where $z=r e^{i\theta}$, $r>e^{-d/c}$. For $n>1$,
we obtain what appears to be a system of $p$ black holes, $p\leq n$ being
the number of connected components of the horizon $X(x,y)=0$, the total
horizon perimeter being $2\pi\sum_{i=1}^n c_i$. This solution may be
maximally extended by taking two identical copies of the multiply connected
$X>0$ region, which generalize the two regions $I$ of the Rindler space-time,
and connecting the corresponding $p$ horizon components via $p$ two-sided
bridges made of two copies (past and future) of a region of type $II$.

However, a serious problem with the above construction is that the metric
(\ref{4}) has $n-1$ conical singularities associated with the zeroes of
$Z'(z)$. As conical singularities correspond to point particles, we must
require for consistency \cite{r5} that these follow geodesics of the
multi-black-hole space-time. Now a point particle at rest in the geometry
(\ref{3}) feels a static gravitational field
\begin{equation} \label{6}
-\Gamma_{00}^X=-\,\gamma^2X
\end{equation}
which vanishes only on the horizon $X=0$, and the configurations such that
the zeroes of $Z'(z)$ sit on the horizon are obviously rather special.
Of course, the problem is evaded for those zeroes of $Z'(z)$ which lie
behind the horizon and do not belong to the multi-black-hole space-time
(the regions $X<0$ are cut out and replaced by connecting bridges). However,
for all the zeroes of $Z'(z)$ to lie behind the horizon this must be simply
connected ($p=1$), in which case the space-time geometry reduces to that
of the original Rindler cylinder (if the restriction to positive $c_i$ is
lifted, then static multi-black-hole space-times with all the zeroes of
$Z'(z)$ lying behind the horizon are possible, with several lines at
spatial infinity). The conclusion is that the previously discussed static
multi-black-hole solution is unconsistent. However, the preceding analysis
hints strongly towards a dynamical solution. Consider for instance the map
\begin{equation} \label{7}
Z=c\ln(z^2-a^2)+d
\end{equation}
($c>0$); in the case $c\ln|a|^2+d>0$ this leads to a two-black-hole
`solution' with an unphysical conical singularity located at the
`center of mass' $z=0$. The gravitational field (\ref{6}) acting on this
singularity, which tends to reduce $X(0)$ and thus the separation $2|a|$
between the black hole centers, pulls the two black holes together until
they merge in a single black hole for $c\ln|a|^2+d=0$.

To translate this picture into an exact solution, we must introduce a time
dependence in the multi-black-hole solution. This can be done by generalizing
the conformal map $Z=Z(z)$ to the time-dependent map
\begin{equation} \label{8}
Z=Z(z,t)
\end{equation}
which leads from the static flat metric (\ref{3}) to a dynamical flat metric.
As we want to describe a system of moving black holes, we shall assume
$Z(z,t)$ to be given by (\ref{5}), where the positions $a_i$ of the centers are
now time dependent, leading to the metric
\begin{equation} \label{9}
ds^2=(\gamma^2X^2-|A|^2)dt^2+(\bar{A}Z'dz+A\bar{Z}'d\bar{z})dt-
|Z'^2|dzd\bar{z},
\end{equation}
where
\begin{equation} \label{10}
A(z,t)=\sum_{i=1}^n\frac{\displaystyle c_i\partial_0a_i(t)}
{\displaystyle z-a_i(t)}
\end{equation}
(such a transformation was previously used by Letelier and Gal'tsov \cite{r9}
to construct multiple moving cosmic strings). The metric (\ref{9}) has again
a horizon at $X(z,t)=0$, and $n-1$ conical singularities following the
worldlines $z_\alpha(t)$ which solve the equation $Z'(z_\alpha,t)=0$.
For consistency, these worldlines must obey the geodesic equations
\begin{equation} \label{11}
\ddot{x}_\alpha^\mu+\Gamma_{\nu\rho}^\mu(x_\alpha)\dot{x}_\alpha^\nu
\dot{x}_\alpha^\rho=0,
\end{equation}
where $x_\alpha^0\equiv t$ for all $\alpha$, and $\dot{}\equiv
d/d\sigma_\alpha$, $\sigma_\alpha$ being the affine parameter on the
$\alpha$th geodesic. Eliminating the $\sigma_\alpha$ in favour of coordinate
time $t$, we are left with a system of $2(n-1)$ second-order differential
equations for the $2n$ unknowns ($a_i(t)$, $\bar{a}_i(t)$). The remaining
two-fold arbitrariness is of course due to the possibility of arbitrary
global time-dependent translations $z\rightarrow z+w(t)$; if we choose for
instance the origin of the complex $z$-plane to coincide with the
`center of mass' of the multi-black-hole system, then the relative dynamics
of the system are fully determined by integrating the consistency equations
(\ref{11}) with appropriate initial conditions.

We consider in more detail the symmetrical two-black-hole system (\ref{7})
with fixed conical singularity $z=0$. The two-body problem may be reduced to
that of the motion of one `black hole' relative to the fixed point
$\zeta=0$ by the transformation $z^2=\zeta$, which transforms (\ref{7}) to
\begin{equation}\label{12}
Z=c\ln(\zeta-\alpha(t))+d
\end{equation}
with $\alpha=a^2$. Then, this last motion may be transformed, by the global
time-dependent translation $\zeta=\psi+\alpha(t)$, into that of a point
particle following the worldline $\psi+\alpha(t)=0$ relative to the fixed
`black hole' (the background Rindler space-time)
\begin{equation} \label{13}
Z=c\ln\psi+d.
\end{equation}
These successive coordinate transformations mapping the geodesic $z=0$
into a geodesic, it follows that the motion of the two black-hole
centers is given by $\pm a=(-\psi)^{1/2}$, where $\psi=\psi(t)$ is a
geodesic of the Rindler cylinder metric (\ref{13}). These geodesics
may easily be derived from those of Minkowski space-time by the Rindler
transformation. Typical timelike and null geodesics are shown in Fig. 1.
In the timelike case, as the conical singularity crosses the horizon from
region $I$\/ into region $II$, a single black hole bifurcates into two
black holes which separate to a finite distance and merge again (the conical
singularity falls back behind the horizon) after an infinite coordinate time.
The global structure of the maximal extension of this space-time is
schematized in Fig. 3; because of its multiply connected topology, this
dynamical solution is clearly not equivalent to a stationary solution with
point singularities. In the lightlike case (Fig. 4) the two black holes,
infinitely separated at $t=-\infty$, fall upon each other and merge, again
after an infinite coordinate time; the time-reversed evolution is equally
possible. In all cases, the total horizon perimeter $4\pi c$ is a constant
of the motion.

We now sketch how the above construction may be generalized to the case of
$\Lambda<0$ cosmological gravity (fuller details shall be given elsewhere).
In the case $J=0$, multi-black-holes may similarly be obtained from the
one-black-hole (\ref{2}) by the time-dependent conformal map $X+iY=Z(z,t)$,
where now $X$ varies between $ml\pi$ (the horizon) and $(m+1/2)l\pi$ (the
line at spatial infinity, which may also be multiply connected) for a given
integer $m$, and the functions $a_i(t))$ in (\ref{5}) are determined by
the condition that the zeroes of $Z'(z,t)$ follow geodesics. The resulting
dynamical picture differs from that of the $\Lambda=0$ case in several
respects. All timelike geodesics of the $\Lambda<0$ black hole cross
periodically the horizon, leading to a pulsating system of two black holes
periodically merging and coming again apart. A class of lightlike geodesics
yield solutions describing the scattering of two black holes, with a
one-black-hole intermediate state. Finally, we can now observe splitting and
merging, not only of black holes, but also of universes (if we define a
universe as a connected component of a spatial section of the full
space-time).

In the general case $J\neq0$ ($J^2\le M^2l^2$), we can write the
one-black-hole
solution (\ref{1}) as
\begin{equation} \label{14}
ds^2=h^2(\nu dt+\frac{J}{2ch^2}dY)^2
-\frac{l^2}{c^2}(h^2+M
+\frac{J^2}{4l^2h^2})(dX^2+dY^2),
\end{equation}
where $h^2=r^2/l^2-M$ is related to $X$ by
\begin{equation} \label{15}
\frac{dh}{dX}=c^{-1}[h^2+M+\frac{J^2}{4l^2h^2}],
\end{equation}
and $Y=c\,\theta$. The construction then proceeds as before, except that the
solution $Z(z,t)$ must be analytically continued beyond $h^2=0$ to the
largest, negative root $h_+^2$ of the right-hand side of (\ref{15}) (the
outer horizon). Particularly interesting is the extreme case $J^2=M^2l^2$,
in which we would expect that the gravitational attraction and the centrifugal
repulsion may balance, resulting in stationary solutions. Indeed, we can show
that the lines $r=a_0$, $d\theta=(\nu/l)dt$ are lightlike geodesics for
arbitrary $a_0$, corresponding to stationary systems of two black holes
orbiting around the conical singularity at the constant angular velocity
$\nu/2l$.

We have studied the classical dynamics of black holes in three-dimensional
cosmological gravity. Limiting cases of special interest are pure gravity
($\Lambda=0$) where, despite the fact that space-time is (almost everywhere)
flat, we have obtained dynamical systems of freely falling black holes and
conical singularities with non trivial topology, and extreme black holes
($J^2=-M^2/\Lambda$), which may interact together with conical singularities
to form stationary planetary systems.

\bigskip
\noindent {\large\bf Acknowledgment}

I wish to thank Bernard Linet for a critical reading of the manuscript.

\newpage
\noindent {\Large\bf Figure captions}
\begin{description}
\item[Fig.1:]
Penrose diagrams for the $\Lambda=0$ black hole (Rindler cylinder),
with a timelike geodesic (dashed line) and a lightlike geodesic (wavy line).
\item[Fig.2:]
Penrose diagram for the black-hole space-time generated by two mirror-symmetric
rings (heavy vertical worldlines). The exterior regions are truncated
conical space-times, while the interior region is a truncated Rindler cylinder.
\item[Fig.3:]
Penrose diagram for the timelike two-black-hole space-time with
$Y=c\pi$ ($\Lambda=0$). The mirror-symmetric conical singularities are
shown as heavy worldlines in the two regions $I$ (their dashed analytic
extensions into the regions $II$ are not associated with conical
singularities).
The double lines result from the superposition, induced by the map $z^2=\zeta$,
of the two disjoint horizon components.
\item[Fig.4:]
Penrose diagram for the lightlike two-black-hole space-time with
$Y=c\pi$ ($\Lambda=0$).
\end{description}
\end{document}